\documentclass[prl,twocolumn,amsmath,amssymb,superscriptaddress]{revtex4}
\usepackage{tipa}
\usepackage{amssymb}
\usepackage{subfigure}
\usepackage{graphicx}
\usepackage{epstopdf}
\usepackage{dcolumn}
\usepackage{bm}
\usepackage{float}
\usepackage{color}
\begin{document}
	
	\preprint{}
	\title{Conformal Quantum Geometry Induced by Entanglement}
	
	\author{Minggui Liang}\thanks{These authors contributed equally to this work.}
	\affiliation{Laboratory for Spin Photonics, School of Physics and Electronics, Hunan University, Changsha 410082, China}
	
	\author{Yan Wang}\thanks{These authors contributed equally to this work.}
	\affiliation{Laboratory for Spin Photonics, School of Physics and Electronics, Hunan University, Changsha 410082, China}
	
	\author{Shuangchun Wen}
	\affiliation{Laboratory for Spin Photonics, School of Physics and Electronics, Hunan University, Changsha 410082, China}
	
	\author{Hailu Luo}\email{hailuluo@hnu.edu.cn}
	\affiliation{Laboratory for Spin Photonics, School of Physics and Electronics, Hunan University, Changsha 410082, China}
	
	\date{\today}
	
	\begin{abstract}
		Quantum metrology beyond the classical limit has relied on high-fidelity multipartite entanglement, which raises the quantum Fisher information~(QFI) but is fragile to loss and noise, while the geometric phase is robust yet has not been linked to sensitivity through a single control parameter. Here we show that entanglement induces a conformal quantum geometry on the Poincar\'e-sphere paths of a polarization-entangled photon pair, in which the Fubini--Study metric and the Berry curvature are governed by one concurrence-set factor. Line elements scale with this factor, and the QFI and geometric phase, which derive from the metric and curvature, scale with its square. In the weak-entanglement, near-orthogonal limit the amplification approaches $2/C$, paid for by a lower success probability, and information is redistributed rather than created. At $99\%$ visibility the conditional contrast exceeds $0.98$, and the saturated gain obeys the finite bound $V^2/(1-V^2)$, reached only in the rare-event limit. The unit-gain threshold sits at $V=1/\sqrt{2}$, the same visibility at which the Werner state first violates the CHSH inequality, so metrological gain and Bell nonlocality emerge simultaneously. Weak-value amplification enters as a limiting case, and the same conformal quantum geometry unifies robust sensing with nonlocal phase engineering.
	\end{abstract}
	
	\maketitle
	
	Accurate measurement of physical quantities underpins the natural sciences. Quantum metrology shows that nonclassical resources such as entanglement can drive the uncertainty of parameter estimation below the classical limit~\cite{Horodecki2009,Afek2010,Israel2014}. Schemes represented by NOON states raise the quantum Fisher information~(QFI) of the entire measurement so that the precision approaches the Heisenberg scaling~\cite{SHuang2025,YZhang2025}, and have been demonstrated across several physical systems~\cite{Graham2022,YChi2022,Kleine2026}. Their advantage rests on high-fidelity multipartite entanglement, and particle loss together with phase diffusion quickly erodes the achievable gain, placing heavy demands on resources and stability~\cite{CHu2025,Cildiroglu2025}. Conditional measurement offers a different route, as it does not add to the total information but concentrates it into a small fraction of selected events. Weak-value amplification~\cite{Hosten2008,Omar2014,WQu2020,LXu2020,JLiu2024} is the representative example. When the postselection is nearly orthogonal, the conditional reading can move outside the eigenvalue range of the observable, but most events are discarded and the operating point is confined to a region of low count rate~\cite{Rath2021,Scandi2025}. These two approaches share a common limitation. Any gain in sensitivity comes at the expense of fragility, either in the entanglement resource that must be kept near perfect or in the statistics drawn from vanishingly rare events. No quantum system is fully isolated, so loss and noise are intrinsic to real measurement rather than avoidable artifacts. The central problem is therefore not simply to enlarge a response, but to realize enhancement that is robust by construction and does not rely on near-perfect entanglement or rare postselected events. Such a mechanism would unite precision with resilience under realistic conditions.  
	
	The geometric phase~\cite{Chiao1986} offers a different angle on this problem. The Pancharatnam--Berry phase acquired by a two-level system evolving along a closed path depends only on the solid angle enclosed by the path, is independent of the dynamical process, and is insensitive to common-mode phase fluctuations, which makes it well suited to robust measurement~\cite{Aharonov1987,Biener2002,Bliokh2008}. These ideas have been extended from cyclic and noncyclic pure-state evolution to general and mixed-state settings and to topological quasiparticles~\cite{YYang2019,JJia2025,JLi2025}, and share the gauge-theoretic origin of the Aharonov--Bohm phase~\cite{Cohen2019}. The ray space of two-level pure states is the complex projective line, diffeomorphic to the Poincar\'e sphere. The QFI and the geometric phase correspond respectively to the Fubini--Study metric and the Berry curvature of this space~\cite{Bengtsson2007,do2016differential}. This structure suggests that if entanglement can set up a geometric mapping between the polarization paths of two photons, a single entanglement parameter might govern both the geometric phase and the QFI. Non-Abelian generalizations and geometric quantum computation have been developed along related lines~\cite{Marrucci2006,Devlin2017}, and the quantum steering of entangled photons has been studied extensively~\cite{Defienne2021,YGao2022,MLiang2025}. Whether continuous steering can be understood as a conformal quantum geometry of the sphere that governs phase amplification and sensitivity enhancement through one parameter, and whether such a geometry remains effective at finite visibility, remain open questions.
	
	\begin{figure*}[t]
		\centerline{\includegraphics[width=\linewidth]{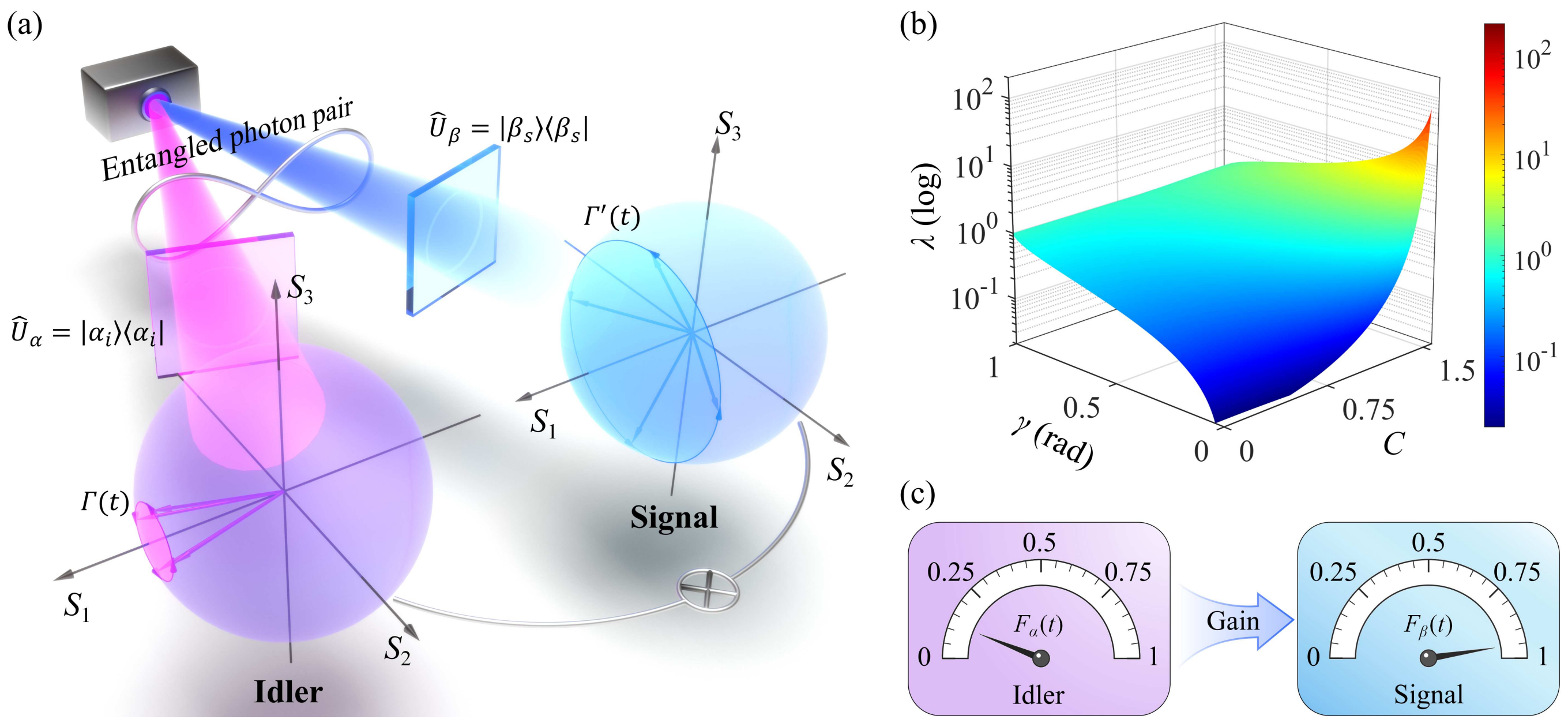}}
		\caption{\label{Fig1} Theoretical framework of the conformal quantum geometry. (a) Conceptual schematic in which an entangled photon pair is split into idler and signal arms, each projected onto a Poincar\'e sphere, and the two outcomes are compared by a coincidence measurement. (b) Three-dimensional surface of $\lambda$ on a logarithmic vertical axis with base-plane contours. (c) Gauge picture of the enhancement, in which the gain raises the low idler Fisher-information reading $F_\alpha(t)$ to a high signal reading $F_\beta(t)$, illustrating how $\lambda$ amplifies the parameter sensitivity.}
	\end{figure*}
	
	In this work we show that entanglement induces a conformal quantum geometry between the Poincar\'e-sphere paths of a polarization-entangled photon pair, in which one conformal factor set by the concurrence governs both the geometric phase and the QFI through the Berry curvature and the Fubini--Study metric. We treat finite visibility with density matrices and show that at $99\%$ visibility the conditional contrast stays above $0.98$, a level readily reached by present polarization-entangled sources~\cite{YZheng2024,XLuo2026,MLiang2026}. An irreducible background-noise contribution keeps the saturated gain finite and equal to the bound $V^2/(1-V^2)$, reached at the rare-event end. A cardioid path that can be synthesized with wave plates and a near-orthogonal circle whose success probability tends to zero examine the conclusion from the high-count-rate and rare-event ends, and weak-value amplification enters the same framework as a limiting case. The same geometry also ties metrological gain to nonlocality, since the visibility where the conditional QFI first exceeds the idler QFI is the visibility at which the Bell inequality is first violated. These results bring phenomena previously separated between Heisenberg metrology, weak measurement, and Bell nonlocality within one conformal quantum geometry. Tuned continuously through the concurrence, this geometric basis supports entanglement-assisted robust sensing, nonlocal phase engineering, and phase estimation under realistic loss and noise.
	
	Consider a polarization-entangled photon pair
	\begin{equation}
		|\Phi\rangle=\cos\kappa\,|H_sH_i\rangle+\sin\kappa\,e^{i\delta}|V_sV_i\rangle ,
	\end{equation}
	where $H$ and $V$ denote horizontal and vertical polarization and $s$, $i$ label the two photons. We write the concurrence as $C = \sin 2\kappa$ and introduce $q \equiv \cos 2\kappa$. These two variables satisfy $C^2 + q^2 = 1$, with the sign of $q$ selecting which polarization the conditional state is biased toward. Projecting the idler onto $|\alpha_i\rangle=\cos\gamma|H_i\rangle+\sin\gamma\,e^{i\varphi}|V_i\rangle$, the conditional state of the signal is
	\begin{equation}
		|\beta_s\rangle
		=\cos\kappa\cos\gamma\,|H_s\rangle
		+\sin\kappa\sin\gamma\,e^{i(\delta-\varphi)}|V_s\rangle .
	\end{equation}
	As the idler projection direction varies continuously, the signal is steered through successive polarizations, the manifestation of quantum steering in the continuous polarization space~\cite{YGao2022,MLiang2025}. To expose the geometric relation between the two paths, we write the Stokes vectors in spherical coordinates and differentiate the conditional state (see Appendices~\ref{AA} and \ref{AB} for more details). The increments of the polar and azimuthal angles obey
	\begin{equation}
		d\theta'=\lambda\,d\theta,\quad d\varphi'=-d\varphi,\quad
		\lambda(\gamma,C)=\frac{C}{1+q\cos 2\gamma}.
	\end{equation}
	The two angular increments are linked by the same position-dependent factor, and the local included angle is preserved. Therefore, the mapping is conformal on the sphere, with $\lambda$ being its local Jacobian. Equivalently, the metric on the mapped sphere equals $\lambda^2$ times the original metric at every point. All local angles and intersections, as well as the topology of the idler path, are therefore retained and no shear distortion appears. The position dependence of $\lambda$ only rescales a finite path nonuniformly and changes its curvature, which is an isotropic scaling rather than an angular distortion. Consequently, only the local scale and handedness of the path are altered. We use conformal in this metric sense throughout, namely a pointwise positive rescaling $g'=\lambda^2g$ that preserves local angles. 
	
		\begin{figure}[t]
		\centerline{\includegraphics[width=\linewidth]{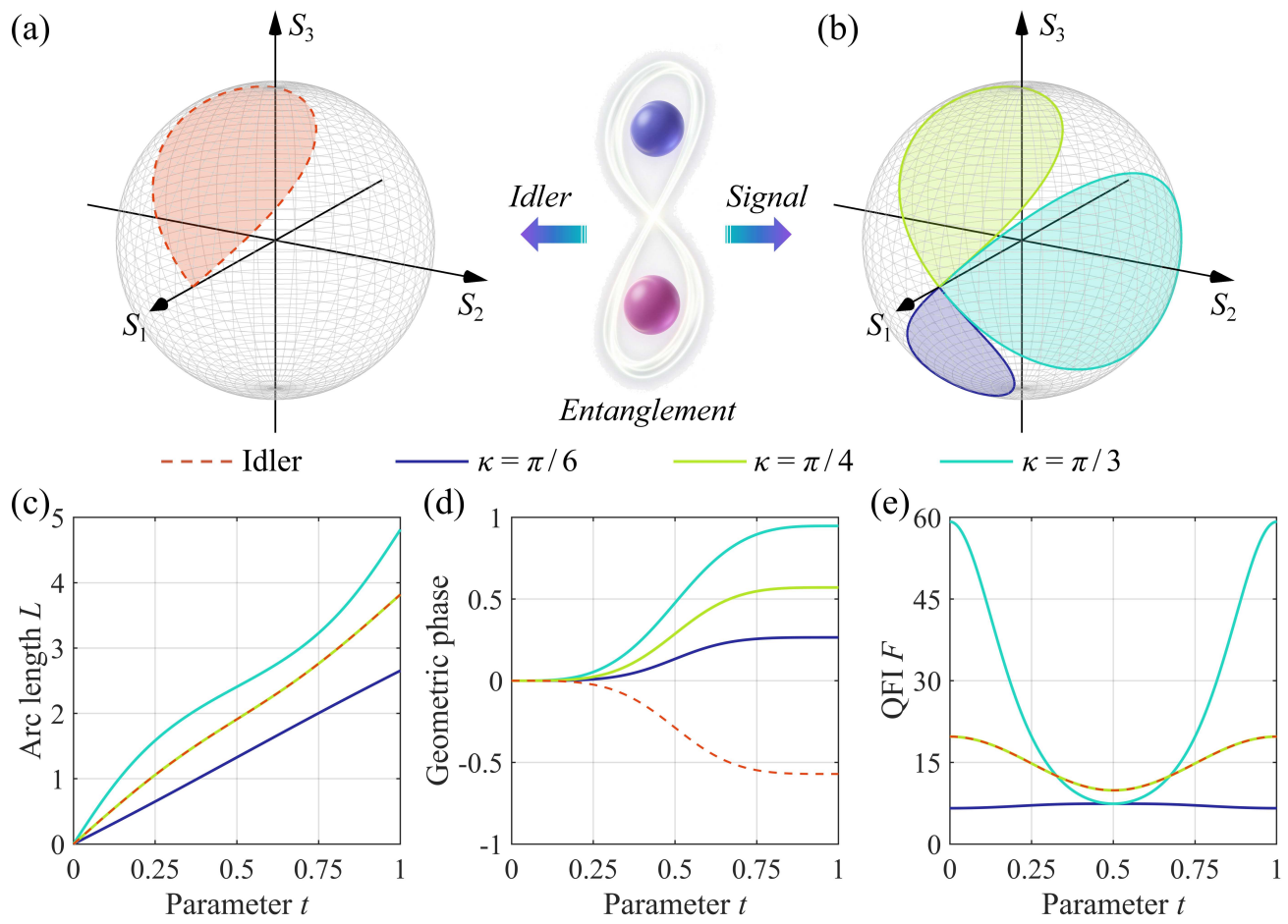}}
		\caption{\label{Fig2} Global and quantitative characterization of the conformal quantum geometry. (a) Poincar\'e sphere of the idler path. (b) Poincar\'e sphere of the signal path. (c) Accumulated arc length. (d) Geometric phase. (e) Local QFI.}
	\end{figure}
	
	As shown in Fig.~\ref{Fig1}(a), the entangled photon pair is split into idler and signal arms, the projectors $\hat U_\alpha=|\alpha_i\rangle\langle\alpha_i|$ and $\hat U_\beta=|\beta_s\rangle\langle\beta_s|$ select a polarization on each Poincar\'e sphere, and a coincidence measurement correlates the two. As the idler projection state evolves with the parameter $t$ along the closed path $\Gamma(t)$ on its Poincar\'e sphere, the conditional signal polarization traces the corresponding closed path $\Gamma'(t)$, where the ratio of the opening angles of these two paths is governed by the concurrence. Figure~\ref{Fig1} then shows the overall behavior of $\lambda$. In Fig.~\ref{Fig1}(b) the three-dimensional surface is drawn on a logarithmic vertical axis with contours projected onto the base plane. The logarithmic scale keeps the gentle variation in the low-gain region and the steep rise in the high-gain region visible at once. Over most of the parameter space $\lambda$ varies mildly and grows sharply only within a narrow region where the projection angle approaches the boundary, and this boundary narrows as the entanglement weakens. The gauge picture in Fig.~\ref{Fig1}(c) illustrates the same enhancement in a direct way, where the gain moves the low idler Fisher-information reading to a high signal reading, so that the pointer deflection clearly illustrates the amplification of parameter sensitivity by $\lambda$. In the weak-entanglement, near-orthogonal limit,  
	\begin{equation}
		\lambda_{\max}=\frac{C}{1-\sqrt{1-C^2}}
		=\frac{2}{C}-\frac{C}{2}+O(C^3).
	\end{equation}
	This does not imply that weaker entanglement gives a larger advantage. Its geometric origin is that, for weak entanglement, a broad region of the idler sphere is compressed into a small neighborhood of the signal sphere. A small parameter change within that neighborhood corresponds to a wide sweep over the compressed region, producing a large local Jacobian at the boundary, and the more nearly orthogonal the projection, the larger the compressed region. The amplification is paid for by the rarity of the conditional events, whose success probability is
	\begin{equation}
		P=\frac{1}{2}\left(1+q\cos 2\gamma\right).
	\end{equation}
	Comparing the two expressions, the larger $\lambda$ is, the smaller $P$, and the high-gain region coincides with the rare-event region. Weighted by the success probability, the conditional measurement does not exceed the total QFI available without postselection. What the concurrence does is move information from the discarded events to the retained ones, rather than create it.
	
	For a pure state, the QFI describes the distinguishability of states along the parameter direction, while the geometric phase results from the Berry connection accumulated along a closed path, and the two appear unrelated. The pure-state ray space of a two-level system is the complex projective line $\mathbb{CP}^1$, diffeomorphic to the Poincar\'e sphere. The Hopf fibration from the Hilbert sphere $S^3$ to $\mathbb{CP}^1$ carries a natural Berry connection whose curvature determines the geometric phase, and the Fubini--Study metric on the same ray gives the distance between states. In Bloch coordinates the area form of this metric is proportional to the Berry curvature, so distance and curvature are not two independent objects~(see Appendix~\ref{AC} for more details).
	
	\begin{figure}[t]
		\centerline{\includegraphics[width=\linewidth]{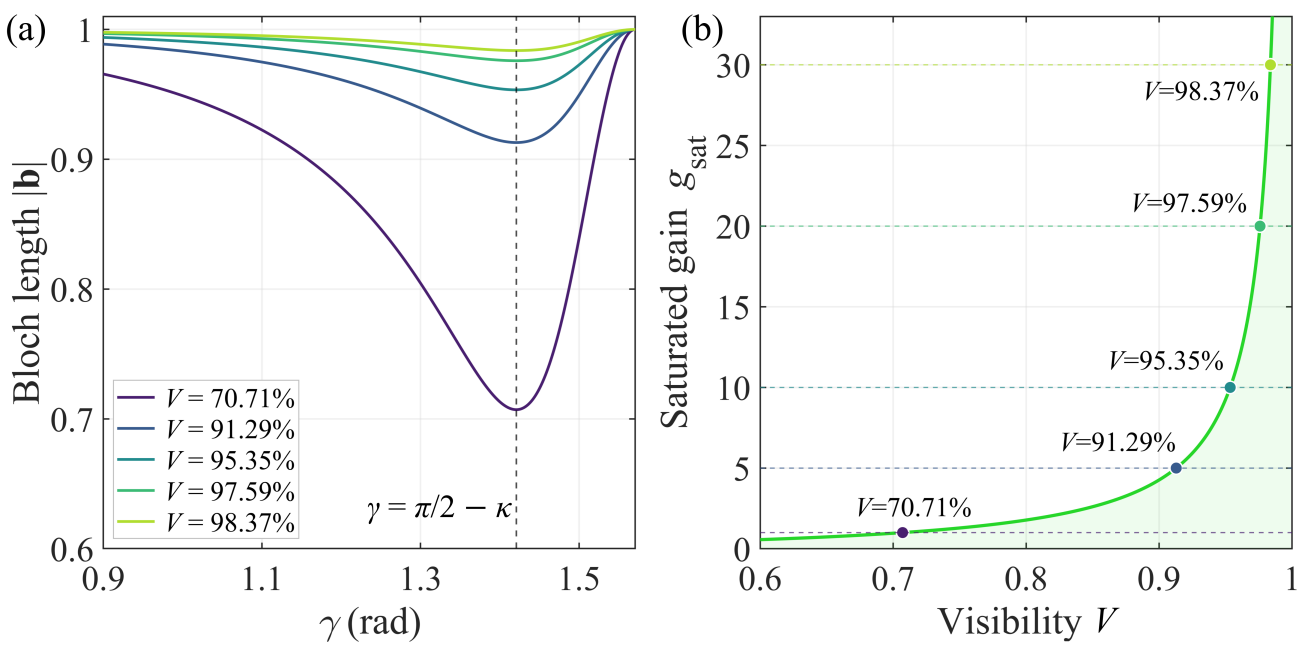}}
		\caption{\label{Fig3} Effectiveness of the enhancement at finite visibility.
			(a) Conditional contrast at the five visibilities corresponding to target gains
			of $1$, $5$, $10$, $20$, and $30$, with the dashed vertical line marking their common minimum at
			$\gamma=\pi/2-\kappa$. (b) Saturated gain, with the visibility each
			target requires marked. Colors are matched across the two panels.}
	\end{figure}
	
	This shared geometry leads directly to the scaling of area and information. From the angular increment one verifies $\sin\theta'=\lambda\sin\theta$, so the spherical area element obeys
	\begin{equation}
		dA_\beta=\lambda^2 dA_\alpha ,
	\end{equation}
	The geometric phase is minus one half of the solid angle, and integrating the scaling of the area element along a closed path gives the amplification of the geometric phase. On the other hand, the QFI of a pure polarization state equals the squared norm of its Stokes velocity, hence
	\begin{equation}
		F_\alpha=\lVert\dot{\mathbf{s}}_\alpha\rVert^2,\qquad
		F_\beta=\lambda^2 F_\alpha .
	\end{equation}
	The factor $\lambda^2$ in the area and information relations appears because both involve the product of two directions, whereas length involves only one. That one concurrence can govern both the geometric phase and the QFI follows from the common origin of the Fubini--Study metric and the curvature. The mapping changes the metric, so distance and information change, and it changes the curvature at the same time, so the solid angle and phase change. The reversal of the azimuthal angle also makes the two paths opposite in handedness and their geometric phases opposite in sign.
	
	\begin{figure*}[t]
		\centerline{\includegraphics[width=\linewidth]{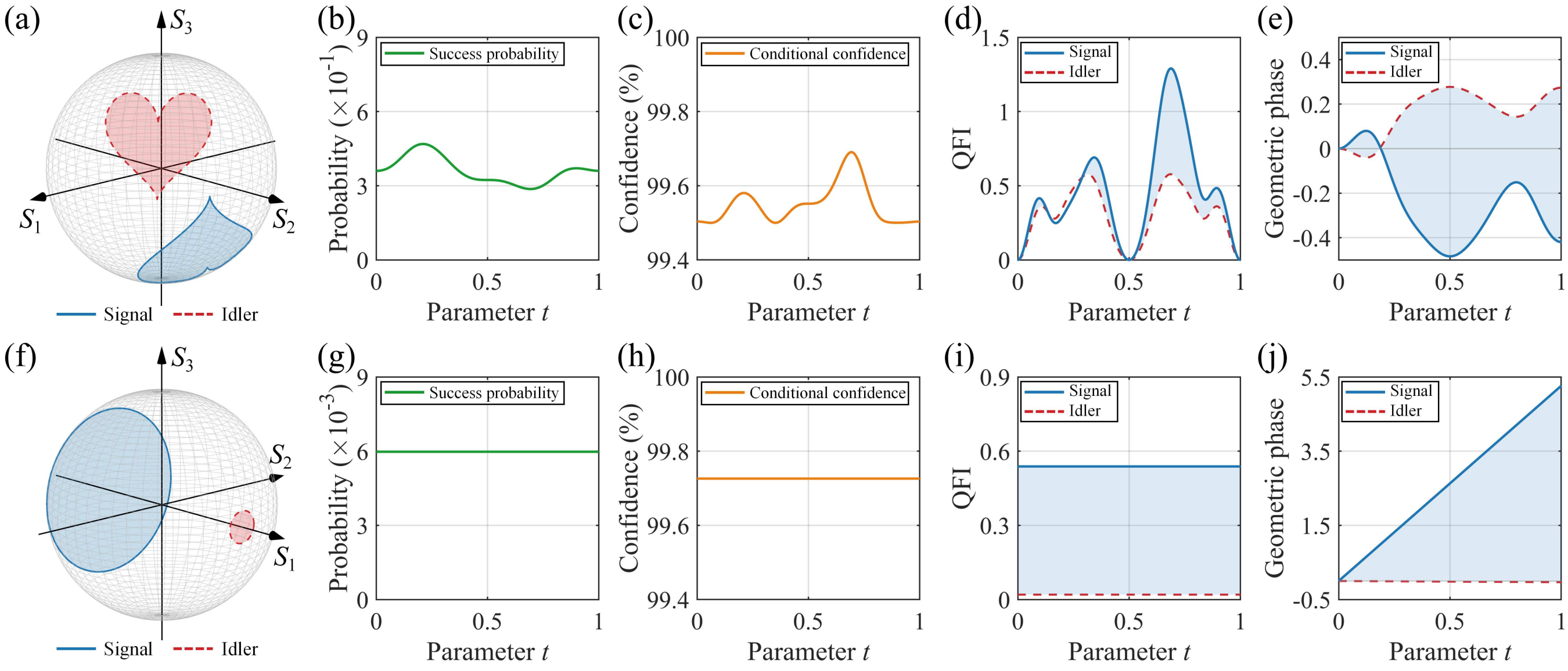}}
		\caption{\label{Fig4} Comparison of the two closed paths, with the cardioid in the top row and the near-orthogonal circle in the bottom row. (a) Idler and signal cardioid paths on one sphere. (b) Cardioid success probability. (c) Cardioid conditional confidence. (d) Cardioid QFI comparison with the enhancement region shaded. (e) Cardioid geometric-phase comparison with the enhancement region shaded. (f) Idler and signal circular paths on one sphere. (g) Circular success probability. (h) Circular conditional confidence. (i) Circular QFI comparison with the enhancement region shaded. (j) Circular geometric-phase comparison with the enhancement region shaded.} 
	\end{figure*}
	
	For a path produced by a quarter-wave plate, Fig.~\ref{Fig2} quantitatively presents the various aspects of these relations. Figures~\ref{Fig2}(a) and \ref{Fig2}(b) draw the idler and signal paths on two separate Poincar\'e spheres. The idler sphere contains a single reference path, while the signal sphere overlays the images at three concurrences. Comparison of the two shows that the path topology remains invariant, while the scale and chirality change. The accumulated arc length in Fig.~\ref{Fig2}(c) shows how different concurrences stretch or compress the same idler path to different total lengths. In the geometric phase of Fig.~\ref{Fig2}(d), each signal curve has the opposite sign to the idler, and the magnitude of the final phase varies with the concurrence. The local QFI in Fig.~\ref{Fig2}(e) varies along this non-geodesic path rather than staying constant. Each signal curve equals the idler value multiplied by $\lambda^2$, enhanced where $\lambda$ exceeds unity and suppressed elsewhere, at positions matching Fig.~\ref{Fig1}. At $\kappa=\pi/4$, $\lambda$ is unity throughout and the map reduces to an orientation-reversing isometry in which the two paths are mirror images on the sphere, so the signal QFI equals the idler value at every point~\cite{YWang2026}.  
	
	The divergence $2/C$ in the weak-entanglement expression assumes that the conditional state is pure. A real entangled source has finite interference visibility, and the conditional state should be represented by a density matrix whose Bloch-vector length is below unity. Consider first the coherence-decay model, which corresponds directly to interferometric measurement. The populations of the entangled state are unchanged and the off-diagonal coherence terms are attenuated by the visibility $V$. After projecting the idler, the conditional density matrix of the signal (its full form is given in Appendix~D) is written as $\varrho_\beta=(1/2)(I+\mathbf{b}\cdot\boldsymbol{\sigma})$, and its Bloch length obeys~\cite{Bengtsson2007}
	\begin{equation}
		|\mathbf{b}|^2
		=1-\frac{(1-V^2)C^2\sin^2 2\gamma}
		{(1+q\cos 2\gamma)^2}.
	\end{equation}
	The correction is small overall. At the practical operating point $V=0.99$, $|\mathbf{b}|>0.98$ and the conditional state lies almost on the sphere. Its Fubini--Study geometry differs from the pure-state case negligibly at almost every point. Writing $1-|\mathbf{b}|^2=\epsilon$, the radial term $(\mathbf{b}\cdot\dot{\mathbf{b}})^2/(1-|\mathbf{b}|^2)$ is of order $\epsilon$, whereas the transverse term $|\dot{\mathbf{b}}|^2$ carries the leading $\lambda^2$ scaling. Along the cardioid path at $V=0.99$, $|\mathbf{b}|\geq 0.99$ and the radial term stays below $4\times10^{-3}$ of the total QFI, so the gain is retained almost in full.
	
	Experiments also contain signal-independent backgrounds from multiphoton events, dark counts, and environmental decoherence, which give nonzero counts even at ideal orthogonality. Including this worst-case scenario, we take a Werner-type state~\cite{Werner1989}
	\begin{equation}
		\rho_W=V|\Phi\rangle\langle\Phi|+(1-V)\frac{I}{4}.
	\end{equation}
	Its conditional success probability is $P_W=VP+(1-V)/2$. At the endpoint where the pure-state model gives $P\to0$, the white noise supplies a nonzero floor, the normalization singularity disappears, and the amplification becomes finite. The mixed-state QFI contains a radial suppression term that grows as the Bloch length shortens [Eq.~(\ref{eqd6})]. Maximizing the conditional QFI over the operating point and the path direction gives the strict saturated gain,
	\begin{equation}
		g_{\max}
		=\frac{V^2}{1-V^2}.
	\end{equation}
	
	This bound is reached only in the rare-event limit. Figure~\ref{Fig3} gives the finite-visibility behavior from two complementary viewpoints, using the same five visibilities and matched colors in both panels. Figure~\ref{Fig3}(b) reads the saturated-gain curve in the inverse direction. The five dashed lines mark target gains of $1$, $5$, $10$, $20$, and $30$, and their intersections with the curve give the visibility each target requires, respectively $70.71\%$, $91.29\%$, $95.35\%$, $97.59\%$, and $98.37\%$. Figure~\ref{Fig3}(a) plots the conditional contrast at exactly these five visibilities. The four higher-visibility curves remain near or above $0.913$, while the lowest visibility dips to about $1/\sqrt{2}$. All five curves reach their minimum at the same projection angle $\gamma_{\min}=\pi/2-\kappa$, marked by the dashed line. The visibility sets the depth of the dip but not its position, so reducing the visibility shortens the Bloch vector without shifting the location of its minimum. The lowest of these visibilities, $V=1/\sqrt{2}$, has a distinct physical significance~\cite{Clauser1969}. The maximal CHSH correlation of the Werner state is $2\sqrt{2}V$, which exceeds the Bell bound of two precisely when $V>1/\sqrt{2}$. The condition $g>1$ therefore holds over exactly the same visibility range as a Bell violation, and the metrological advantage opens at the same threshold as the nonlocal character of the state. In both the $D/A$ and $H/V$ bases, visibilities near $99\%$ are readily reached by present polarization-entangled sources~\cite{MLiang2026}, so under the coherence-decay model and the Werner noise-floor model the enhancement remains effective at this visibility. The large gain at the upper end is confined to the singularity where the success probability tends to zero and counts cannot accumulate, and it does not constrain a real measurement.
	
	The local relations above need to be tested on concrete closed paths. We choose two classes, whose parametrization and wave-plate realization are given in Appendix~E. The first is the cardioid path, along which the idler evolves on a closed curve of nontrivial curvature, taking $\kappa=\pi/3$ and $V=0.99$. The second is the near-orthogonal circle, which lies in a plane parallel to the $S_2S_3$ plane so that the $S_1$ component of the idler stays constant. A small circle near the H pole is taken and realized with weak entanglement $\kappa=0.49\pi$, driving the success probability toward zero.

	To validate these local relations and the trade-off between success probability and geometric amplification, we examine two concrete closed paths that probe complementary regimes. One is a cardioid path operating at high count rates, and the other is a near-orthogonal circular path approaching the rare-event limit. Figure~\ref{Fig4} presents this comparison, with the top and bottom rows corresponding to the cardioid and the circular path, respectively. For the cardioid path, the idler and signal trajectories on the Poincaré sphere exhibit similar shapes but opposite handedness [Fig.~\ref{Fig4}(a)]~\cite{YWang2025,YWang2026}. This directly confirms the topology-preserving and orientation-reversing nature of the conformal mapping. At this high-count-rate regime, the success probability remains between $29\%$ and $47\%$ [Fig.~\ref{Fig4}(b)], and the conditional confidence stays close to unity [Fig.~\ref{Fig4}(c)]. These results indicate that the conditional state remains nearly pure. Even without entering the rare-event limit, the geometric mapping yields a substantial enhancement. The conditional QFI peaks at roughly $2.24$ times the idler value [Fig.~\ref{Fig4}(d)], and the geometric phases for the signal and idler exhibit opposite signs with clear amplification [Fig.~\ref{Fig4}(e)]. These features quantitatively corroborate the $\lambda^2$ scaling relations derived from the area element and the QFI definitions.
	
	The near-orthogonal circular path, by contrast, pushes the system into the extreme weak-entanglement limit. The idler evolves along a small loop near the $H$ pole, while its signal image is significantly pulled toward the center of the sphere and enlarged [Fig.~\ref{Fig4}(f)]. The success probability drops dramatically to the rare-event region of $0.6\%$ [Fig.~\ref{Fig4}(g)], yet the conditional confidence remains near unity [Fig.~\ref{Fig4}(h)]. This extreme rarity acts as the physical source of the enormous geometric amplification. The signal QFI exceeds the idler value by a factor of $27$ [Fig.~\ref{Fig4}(i)], and the accumulated geometric phase reaches $5.26$ which wraps to a principal value of approximately $-1.02$ in an experiment [Fig.~\ref{Fig4}(j)]. Consequently, these two cases reveal a central underlying mechanism. The same conformal quantum geometry unifies high-count-rate deterministic sensing with extreme weak-value-like amplification. The concurrence continuously tunes the trade-off between success probability and sensitivity across both regimes.

	This comparison helps to delineate the relation between the present mechanism and conventional weak measurement. The near-orthogonal circle is structurally identical to weak-value amplification. Weak entanglement guarantees that the conditional response can be linearized and plays the part of the weak coupling, while the nearly orthogonal polarization projection is the postselection. Its small success probability is precisely the source of the amplification, and amplifying the conditional reading does not add to the probability-weighted total information. In this sense, the near-orthogonal postselection described by weak-value amplification is the formal limiting case in which the postselection probability tends to zero. As the idler loop approaches the $H$ pole, $P\to0$ while the conditional gain grows as $1/P$, exactly as the weak value $A_w$ does, and the product $P\,g$ approaches a finite constant of order unity, which expresses conservation of the probability-weighted information. The difference lies in the viewpoint and the range of control. Weak measurement attributes the amplification to the algebraic object of the weak value and usually pushes the operating point to a very low success probability. Here the same phenomenon is placed within the same conformal quantum geometry, so the amplification is tuned continuously by the concurrence and the projection angle. The cardioid path shows that, without entering the region where the success probability tends to zero, deterministic geometric-phase amplification and QFI enhancement can be obtained at high count rates. Entanglement-assisted weak-value amplification and the preservation of QFI under postselection fit the same picture, as do the single-photon nonlinear effects reached through weak measurement. The postselection technique, previously treated in isolation within weak measurement, is thereby integrated into a geometric framework tunable continuously through the concurrence.
	
	In summary, we have shown that entanglement induces a conformal quantum geometry in which the control of the geometric phase by quantum correlation and the enhancement of metrological sensitivity are the same physical process. The Fubini--Study metric and the Berry curvature share an origin, so path lengths scale with the conformal factor $\lambda$ while the geometric phase and the QFI scale with its square. At the weak-entanglement boundary, the conditional gain approaches $2/C$, reflecting an information redistribution between channels rather than its creation. Our density-matrix treatment keeps the conditional contrast above $0.98$ at $99\%$ visibility, and the background-noise floor limits the saturated gain to the strict bound $V^2/(1-V^2)$. A key feature is that the unit-gain point coincides with the CHSH violation threshold at $V=1/\sqrt2$, so the metrological advantage and Bell nonlocality open over exactly the same visibility range. The cardioid and near-orthogonal circle validate these predictions from the high-count-rate and rare-event ends, respectively, with weak-value amplification emerging as a formal limiting case. By tying the robust geometric phase and the sensitive QFI to one concurrence-tuned factor, this framework provides a concrete geometric basis for entanglement-assisted robust sensing and nonlocal phase engineering.
	
	Earlier routes to entanglement-enhanced metrology have often treated the global QFI enhancement via high-fidelity multipartite entanglement and the conditional amplification via near-orthogonal postselection as two isolated strategies. The spin-orbit geometry underlying the spin Hall effect of light and related flat-optics phenomena shares the same geometric language but does not tie the phase to a metrological gain~\cite{Aiello2009,Bliokh2015}, while squeezed-state, quantum-sensing, and quantum-rate arguments address the noise and the maximal rate of evolution from a complementary direction~\cite{Rath2021,YZheng2024,Scandi2025,CHu2025}. Our framework unifies the two strategies by identifying the differential geometry of quantum states directly with the entanglement-induced map. Crucially, the conformal factor is set by the concurrence rather than by a specific two-level structure, which allows the construction to carry over to higher-dimensional and mixed-entanglement states. This provides a rigorous and generalizable design principle for quantum measurements that maintain their metrological advantage under the loss and noise unavoidable in real experiments.
	
	\vspace{\baselineskip}
	
	This work was supported by the National Natural Science Foundation of China (Grants No.~12174097).
	
	\appendix
	\setcounter{secnumdepth}{1}
	
	\section{Explicit form of the Stokes vectors}
	\label{AA}
	For a general polarization state $X|H\rangle+Y|V\rangle$, the normalized Stokes components are defined by
	\begin{equation}
		s_1=\frac{|X|^2-|Y|^2}{|X|^2+|Y|^2},\quad
		s_2=\frac{2\,\mathrm{Re}[X^*Y]}{|X|^2+|Y|^2},\quad
		s_3=\frac{2\,\mathrm{Im}[X^*Y]}{|X|^2+|Y|^2},
	\end{equation}
	which obey $s_1^2+s_2^2+s_3^2=1$. Here $s_1=+1$ corresponds to $H$ and $s_1=-1$ to $V$. For left circular polarization $L=(H+iV)/\sqrt2$, substitution gives $s_3=+1$, the north pole, while right circular polarization corresponds to $s_3=-1$.
	
	The Stokes vector of the idler projection state is
	\begin{equation}
		\mathbf{s}_\alpha=
		[\cos 2\gamma,\ \sin 2\gamma\cos\varphi,\ \sin 2\gamma\sin\varphi]^T.
	\end{equation}
	For the conditional state, set $a=\cos\kappa\cos\gamma$ and $b=\sin\kappa\sin\gamma$. The normalization denominator is $a^2+b^2=(1+q\cos 2\gamma)/2$. Substituting component by component gives
	\begin{equation}
		\mathbf{s}_\beta=
		\frac{1}{1+q\cos 2\gamma}
		\begin{bmatrix}
			q+\cos 2\gamma\\
			C\sin 2\gamma\cos(\delta-\varphi)\\
			C\sin 2\gamma\sin(\delta-\varphi)
		\end{bmatrix}.
	\end{equation}
	The normalization factor depends on $\gamma$, showing that the mapping is not a rigid rotation.
	
	\section{Derivation of the conformal mapping}
	\label{AB}
	Introduce an effective polar angle $\theta'$ and azimuthal angle $\varphi'$, set $\theta=2\gamma$, and take
	\begin{equation}
		\varphi'=\delta-\varphi,\qquad
		\tan\frac{\theta'}2=\frac{C\tan\gamma}{1+q}
		=\frac{C}{1+q}\tan\frac{\theta}{2}.
	\end{equation}
	From $C^2=(1-q)(1+q)$ one has $C/(1+q)=(1-q)/C$. Differentiating,
	\begin{equation}
		\frac12\sec^2\frac{\theta'}2\,d\theta'
		=\frac{C}{1+q}\frac12\sec^2\frac{\theta}{2}\,d\theta.
	\end{equation}
	Using $\sec^2(\theta'/2)=1+\tan^2(\theta'/2)$ to simplify, one obtains
	\begin{equation}
		d\theta'=\frac{C}{1+q\cos\theta}\,d\theta\equiv\lambda\,d\theta,\qquad
		d\varphi'=-d\varphi.
	\end{equation}
	Since the polar and azimuthal increments are related by the same scalar factor, the arc-length element is
	\begin{equation}
		dl_\beta^2=d\theta'^2+\sin^2\theta'\,d\varphi'^2
		=\lambda^2 dl_\alpha^2,
	\end{equation}
	the local included angle is preserved, and the mapping is conformal. Using the half-angle relation above together with $C^2=(1-q)(1+q)$, the denominator obtained by expressing $\sin\theta'$ through $\tan(\theta'/2)$ reduces to $(1+q)(1+q\cos\theta)$, which gives the pointwise identity
	\begin{equation}
		\sin\theta'=\lambda\sin\theta,
	\end{equation}
	This is an identity in $\theta$ and $q$ at every point of the sphere, not a path-averaged or numerical result.
	
	\section{Unification of QFI and the Pancharatnam--Berry phase through Fubini--Study geometry}
	\label{AC}
	The pure-state ray space of a two-level system is $\mathbb{CP}^1$. The Hopf fibration from $S^3$ to $\mathbb{CP}^1$ defines the Berry connection, whose curvature two-form in Bloch coordinates is
	\begin{equation}
		\Omega=\frac12\sin\theta\,d\theta\wedge d\varphi=\frac12\,dA.
	\end{equation}
	The Fubini--Study line element on $\mathbb{CP}^1$ is
	\begin{equation}
		ds_{\mathrm{FS}}^2=\frac14\left(d\theta^2+\sin^2\theta\,d\varphi^2\right),
	\end{equation}
	whose volume form $(1/4)\sin\theta\,d\theta\wedge d\varphi$ differs from the curvature $\Omega$ only by a constant, so the metric and the curvature share an origin. For a pure state depending on a parameter $t$, the QFI is~\cite{BraunsteinCaves1994}
	\begin{equation}
		F=4\left(\langle\dot\psi|\dot\psi\rangle-|\langle\dot\psi|\psi\rangle|^2\right).
	\end{equation}
	In Bloch coordinates, using the Fubini--Study line element, this reduces to
	\begin{equation}
		F=\lVert\dot{\mathbf{s}}\rVert^2
		=\dot\theta^2+\sin^2\theta\,\dot\varphi^2.
	\end{equation}
	The Pancharatnam--Berry phase of a closed path follows from the Berry connection integrated along the path.
	\begin{equation}
		\phi_{\mathrm{PB}}
		=-\frac12\oint(1-\cos\theta)\,d\varphi
		=-\frac12\Omega_{\mathrm{solid}}.
	\end{equation}
	The conformal mapping scales the Fubini--Study geometry by $\lambda^2$. The scaling of the metric gives the information relation in the main text, and the scaling of the curvature gives
	\begin{equation}
		\phi_\beta=-\frac12\oint dA_\beta,
	\end{equation}
	with the corresponding integral taken as $\lambda$ varies along the path. This is the geometric reason the geometric phase and the QFI are governed by the same concurrence.
	\newline\newline
	\section{Full treatment of finite-visibility mixed states}
	
	Under the coherence-decay model, the two-body density matrix is
	\begin{equation}
		\begin{aligned}
			\rho={}&\cos^2\kappa\,|HH\rangle\langle HH|
			+\sin^2\kappa\,|VV\rangle\langle VV|\\
			&+V\sin\kappa\cos\kappa
			\left(e^{i\delta}|HH\rangle\langle VV|+\mathrm{h.c.}\right).
		\end{aligned}
	\end{equation}
	After projecting the idler, the unnormalized conditional density matrix of the signal is
	\begin{widetext}
		\begin{equation}
			\tilde{\varrho}_\beta=
			\begin{bmatrix}
				\cos^2\kappa\cos^2\gamma &
				V\sin\kappa\cos\kappa\sin\gamma\cos\gamma\,e^{i(\delta-\varphi)}\\
				V\sin\kappa\cos\kappa\sin\gamma\cos\gamma\,e^{-i(\delta-\varphi)} &
				\sin^2\kappa\sin^2\gamma
			\end{bmatrix}.
		\end{equation}
	\end{widetext}
	The success probability $P=\cos^2\kappa\cos^2\gamma+\sin^2\kappa\sin^2\gamma$ is independent of $V$. After normalization, write $\varrho_\beta=\tfrac12(I+\mathbf{b}\cdot\boldsymbol{\sigma})$. The Bloch components are
	\begin{equation}
		\begin{aligned}
			&b_1=\frac{\cos^2\kappa\cos^2\gamma-\sin^2\kappa\sin^2\gamma}{P},\\
			&b_\perp=\frac{V\sin 2\kappa\sin 2\gamma}{2P},
		\end{aligned}
	\end{equation}
	with the transverse component decomposed by azimuthal angle as $b_2=b_\perp\cos(\delta-\varphi)$ and $b_3=b_\perp\sin(\delta-\varphi)$. From $|\mathbf{b}|^2=b_1^2+b_\perp^2$, simplification gives
	\begin{equation}
		|\mathbf{b}|^2
		=1-\frac{(1-V^2)C^2\sin^2 2\gamma}{(1+q\cos 2\gamma)^2}.
	\end{equation}
	For the Werner-type state, the conditional success probability is
	\begin{equation}
		P_W=VP+\frac{1-V}{2}.
	\end{equation}
	The QFI of a general mixed state $\rho=(1/2)(I+\mathbf{b}\cdot\boldsymbol{\sigma})$ is~\cite{BraunsteinCaves1994,TothPetz2013}
	\begin{equation}
		\label{eqd6}
		F_{\mathrm{mix}}=|\dot{\mathbf{b}}|^2
		+\frac{(\mathbf{b}\cdot\dot{\mathbf{b}})^2}{1-|\mathbf{b}|^2}.
	\end{equation}
	The first term is the squared Stokes velocity and the second is the radial contribution, which vanishes as $|\mathbf{b}|\to1$ and suppresses information as the Bloch length shortens. Write the idler polarization in its intrinsic spherical coordinates $(\theta,\phi)$ with $\theta=2\gamma$, so that the Fubini--Study line element reads $d\ell_\alpha^2=d\theta^2+\sin^2\theta\,d\phi^2$. The conditional Bloch vector is a function of these coordinates,
	\begin{equation}
		B_1=\frac{V(q+\cos\theta)}{1+Vq\cos\theta},\qquad
		r=\frac{VC\sin\theta}{1+Vq\cos\theta},
	\end{equation}
	where the transverse component points at azimuth $-\phi$. Evaluating the mixed-state QFI as a tensor in these coordinates and comparing it with the idler metric gives the local information gain along any direction independently of how the path is parametrized. Along a meridian the polar component and along a parallel circle the azimuthal component both reduce, in the limit $\theta\to0$ taken at $q=-V$, to the same finite number. The two eigenvalues coincide at this point and the metric is isotropic, so every path direction reaches the identical bound. Along a parallel circle the azimuthal gain has the explicit form
	\begin{equation}
		g_\varphi(\theta,q)=\frac{V^2(1-q^2)}{(1+Vq\cos\theta)^2}.
	\end{equation}
	At fixed $\theta$, differentiating with respect to $q$ gives the stationary point $q=-V\cos\theta$, and the second derivative is negative there, so it is the maximum. Substitution gives $g_\varphi(\theta)=V^2/(1-V^2\cos^2\theta)$, which grows as $\theta$ decreases. The polar gain along a meridian reaches the same expression at the limit. At the endpoint $\theta\to0$, where $q=-V$, The strict saturated gain is
	\begin{equation}
		g_{\max}=\frac{V^2}{1-V^2}.
	\end{equation}
	At the idler $H$ projection and the weak-entanglement point $q=-V$, the conditional success probability $P_W=(1+Vq\cos\theta)/2$ equals $(1-V^2)/2$ and tends to zero as the visibility approaches one. Inverting the bound gives the visibility needed for a target gain $g$,
	\begin{equation}
		V_g=\sqrt{\frac{g}{1+g}},
	\end{equation}
	namely $V\approx70.71\%$, $91.29\%$, $95.35\%$, $97.59\%$, and $98.37\%$ at $g=1,5,10,20,30$, respectively.
	
	For the geometric phase at finite visibility, we use the Pancharatnam--Berry phase of the normalized conditional Bloch direction $\mathbf{d}=\mathbf{b}/|\mathbf{b}|$, equal to minus one half of the solid angle enclosed by $\mathbf{d}$ on the unit sphere. This is a direction-based phase of the conditional state and is the definition used throughout the numerical evaluation, rather than the full Uhlmann parallel-transport phase or the visibility-weighted interferometric phase. At $V=1$ the conditional state is pure and $\mathbf{d}$ is the mapped direction, so the $\lambda^2$ area scaling is exact. At $V<1$, $\mathbf{d}$ differs from the pure-state mapped direction only by terms of order $1-|\mathbf{b}|^2$. On the cardioid at $V=0.99$, the largest angular deviation is below $0.3^\circ$ and the enclosed phase differs from its pure-state value by less than $4\times10^{-4}$, so the area scaling is retained up to this small correction.
	
	\section{Parametrization of the closed paths}
	
	For the cardioid path, the idler Stokes vector is taken in terms of a parameter $t\in[0,2\pi]$ as
	\begin{equation}
		\begin{aligned}
			&x(t)=\frac{B(t)-16\sin^3 t}{36\sqrt2},\\
			&y(t)=\frac{B(t)+16\sin^3 t}{36\sqrt2},\\
			&z(t)=\frac{A(t)}{36},
		\end{aligned}
	\end{equation}
	where
	\begin{equation}
		A(t)=13\cos t-5\cos 2t-2\cos 3t-\cos 4t+17,
	\end{equation}
	\begin{equation}
		B(t)=\sqrt{1296-256\sin^6 t-A(t)^2}.
	\end{equation}
	One can verify $x^2+y^2+z^2=1$ and that the path is strictly closed. The polarization angles are recovered from
	\begin{equation}
		\gamma(t)=\frac12\arccos x(t),\qquad
		\varphi(t)=\arg[y(t)+iz(t)]
	\end{equation}
	and then substituted into Appendix~A to give the signal path.
	
	The near-orthogonal circle lies in the plane $S_1=c$, parallel to the $S_2S_3$ plane, and its intersection with the sphere is a circle.
	\begin{equation}
		\mathbf{s}_\alpha(t)=[c,\ \sqrt{1-c^2}\cos t,\ \sqrt{1-c^2}\sin t]^T.
	\end{equation}
	The corresponding polarization angles are a constant $\gamma=(1/2)\arccos c$ and an azimuth $\varphi=t$. Its success probability is
	\begin{equation}
		P=\frac{1+qc}{2},
	\end{equation}
	and for $q<0$ and $c\to+1$, $P\to0$. The signal image remains a circle in the plane $S_1=c'$, where
	\begin{equation}
		c'=\frac{q+c}{1+qc},
	\end{equation}
	and the local scaling factor is constant,
	\begin{equation}
		\lambda=\frac{C}{1+qc},\qquad
		\frac{F_\beta}{F_\alpha}=\lambda^2.
	\end{equation}
	For weak entanglement and $c=0.99$, $P\approx0.006$ and $\lambda\approx5.25$, consistent with the values of Fig.~\ref{Fig4} in the main text. Both paths can be synthesized in the idler arm by a combination of a quarter-wave plate and a half-wave plate.
	
	\bibliography{refs}
	
\end{document}